\documentclass[sigconf,nonacm]{acmart}

\usepackage{booktabs} 

\newcommand{\includevisio}[2][]{\includegraphics[clip, trim=0.5cm 0.5cm 0.5cm 0.5cm, #1]{#2}} 

\usepackage{array}
\newcommand{\PreserveBackslash}[1]{\let\temp=\\#1\let\\=\temp}
\newcolumntype{C}[1]{>{\PreserveBackslash\centering}p{#1}}
\newcolumntype{R}[1]{>{\PreserveBackslash\raggedleft}p{#1}}
\newcolumntype{L}[1]{>{\PreserveBackslash\raggedright}p{#1}}

\usepackage{todonotes} 
\usepackage{amsmath}
\usepackage{bm}
\usepackage{amsfonts}
\usepackage{amsthm}
\usepackage{multirow}
\usepackage{colortbl}
\usepackage{enumitem}
\usepackage{subcaption}
\usepackage[normalem]{ulem}

\usepackage{xcolor}

\def\eqref#1{equation~\ref{#1}}

\def\1{\bm{1}}

\DeclareMathAlphabet{\mathsfit}{\encodingdefault}{\sfdefault}{m}{sl}
\SetMathAlphabet{\mathsfit}{bold}{\encodingdefault}{\sfdefault}{bx}{n}

\copyrightyear{2020} 
\acmYear{2020} 
\setcopyright{acmlicensed}
\acmConference[WSDM '20]{Proceedings of the 13th ACM International Conference on Web Search and Data Mining}{February 2020}{Houston,Texas}

\begin{document}

\title{TU Wien @ TREC Deep Learning '19 -- Simple~Contextualization~for~Re-ranking}

\author{Sebastian Hofst{\"a}tter}
\affiliation{%
  \institution{TU Wien}
}
\email{s.hofstaetter@tuwien.ac.at}

\author{Markus Zlabinger}
\affiliation{%
  \institution{TU Wien}
}
\email{markus.zlabinger@tuwien.ac.at}

\author{Allan Hanbury}
\affiliation{%
  \institution{TU Wien}
}
\email{hanbury@ifs.tuwien.ac.at}

\begin{abstract}

The usage of neural network models puts multiple objectives in conflict with each other: Ideally we would like to create a neural model that is effective, efficient, and interpretable at the same time. However, in most instances we have to choose which property is most important to us. We used the opportunity of the TREC 2019 Deep Learning track to evaluate the effectiveness of a balanced neural re-ranking approach. We submitted results of the TK (Transformer-Kernel) model: a neural re-ranking model for ad-hoc search using an efficient contextualization mechanism. TK employs a very small number of lightweight Transformer layers to contextualize query and document word embeddings. To score individual term interactions, we use a document-length enhanced kernel-pooling, which enables users to gain insight into the model.  
Our best result for the passage ranking task is: 0.420 MAP, 0.671 nDCG, 0.598 P@10 (TUW19-p3 full). Our best result for the document ranking task is: 0.271 MAP, 0.465 nDCG, 0.730 P@10 (TUW19-d3 re-ranking).

\end{abstract}

\maketitle
\section{Introduction}

Our aim in the TREC 2019 Deep Learning track was to evaluate a neural re-ranking model, which balances efficiency, effectiveness, and interpretability. We submitted runs for both the passage and document ranking tasks of the Deep Learning track. We present the TK (Transformer-Kernel) model -- inspired by the success of the Transformer-based BERT model \cite{devlin2018bert,nogueira2019passage} and the simplicity of KNRM (Kernel-based Neural Ranking Model) \cite{Xiong2017}. TK employs a small number of low-dimensional Transformer layers~\cite{vaswani2017attention} to contextualize query and document word embeddings. TK scores the interactions of the contextualized representations with simple, yet effective soft-histograms based on the kernel-pooling technique \cite{Xiong2017}. Additionally, we enhance kernel-pooling with document length normalization (Section \ref{sec:model}).

The main differences of TK in comparison to BERT are:
\begin{itemize}
    \item TK's contextualization uses fewer and lower dimensional Transformer layers with less attention heads. This makes the query-time inference of TK with 2 layers 40 times faster than BERT-Base with 12 layers.
    \item TK contextualizes query and document sequences independently; each contextualized term is represented by a single vector (available for analysis). BERT operates on a concatenated sequence of the query and the document, entangling the representations in each layer.
    \item The network structure of TK makes it possible to analyze the model for interpretability and further studies. TK has an \textit{information bottleneck} built in, through which all term information is distilled: the query and document term interactions happen in a single match matrix, containing exactly one cosine similarity value for each term pair. BERT on the other hand has a continuous stream of interactions in each layer and each attention head, making a focused analysis unfeasible.
\end{itemize}

The differences of TK to previous kernel-pooling methods are:

\begin{itemize}
    \item KNRM uses only word embeddings, therefore a match does not have context or positional information.
    \item CONV-KNRM \cite{Dai2018} uses a local-contextualization with limited positional information in the form of n-gram learning with CNNs. It cross-matches all n-grams in $n^2$ match matrices, reducing the analyzability.
\end{itemize}

Naturally, better efficiency and a restricted information flow through the network comes at the cost of effectiveness. This brings us to our main research question for our participation at TREC'19: \textit{How effective is our balanced model?} To investigate this question we submitted multiple configurations of the TK model. We evaluate a GloVe embedding vs. a FastText embedding, an ensemble of multiple model instances, and a windowed-kernel-pooling for the longer document ranking. 

In addition to a presentation and discussion of our TREC run results (Section \ref{sec:results}) we showcase the analysis and interpretation capabilities of the TK model. We focus on the scenario in which a user would like to understand, for a given query, why two documents are ranked differently. We start by visualizing the word-level similarities (the interaction features) and then we report a limited number of aggregated intermediate results of important kernels (Section \ref{sec:interpretability}).

We publish the source code of the TK model and various neural re-ranking baselines at \textit{github.com/sebastian-hofstaetter/transformer-kernel-ranking}. The repository contains all pre-processing and evaluation code, as well as clear and documented neural network implementations using PyTorch~\cite{pytorch2017} and AllenNLP~\cite{Gardner2017AllenNLP}.

\begin{figure*}
    \centering
    \includevisio[width=1\textwidth]{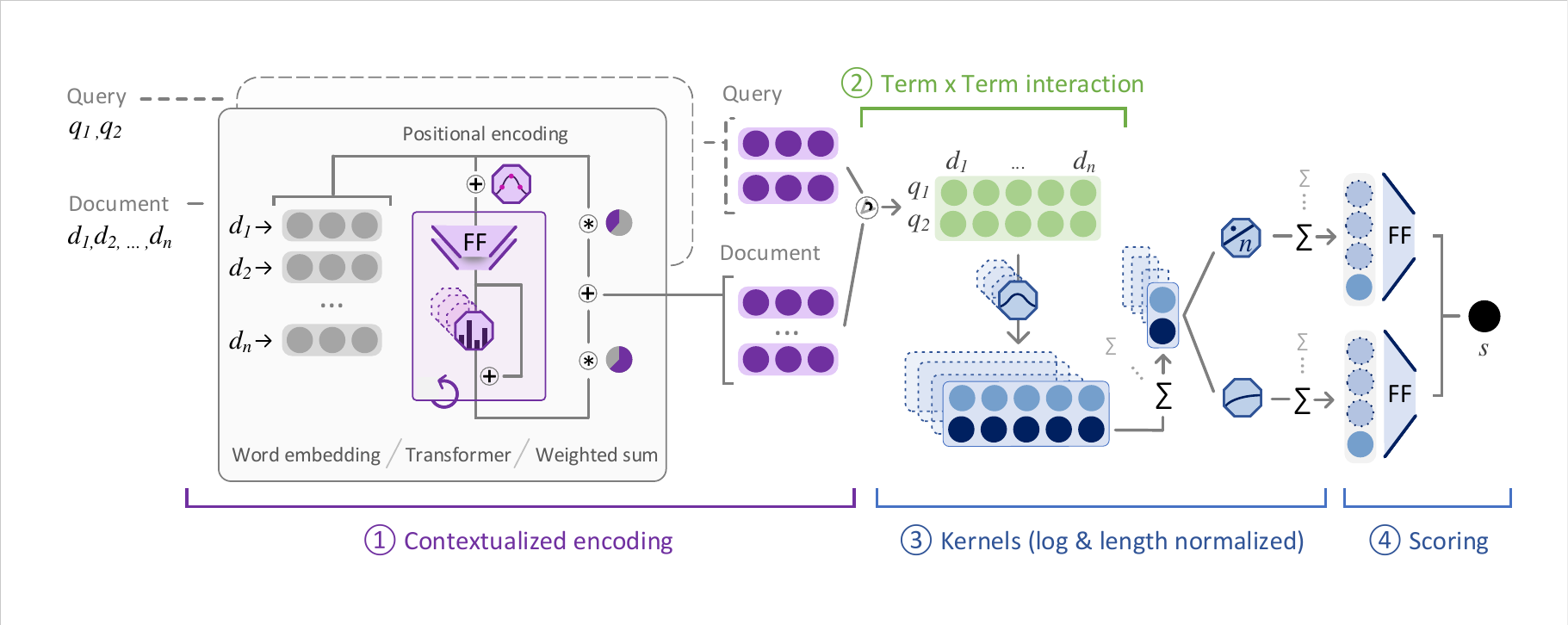}
    \vspace{-0.8cm}
    \caption{The TK model architecture: \raisebox{.5pt}{\textcircled{\raisebox{-.85pt} {1}}}
 We contextualize query and document sequences individually. \raisebox{.5pt}{\textcircled{\raisebox{-.85pt} {2}}} The interaction match-matrix is created with pairwise cosine similarities. \raisebox{.5pt}{\textcircled{\raisebox{-.85pt} {3}}} Each kernel creates a new feature matrix. Then, the document dimension is summed and we normalize each query-term feature by logarithm and document length. \raisebox{.5pt}{\textcircled{\raisebox{-.85pt} {4}}} We combine log- and length-normalized scores to form the final result score.}
    \label{fig:tk_model_architecture}
\end{figure*}
\section{TK: Transformer-Kernel Model}
\label{sec:model}

In this section, we present TK, our Transformer-Kernel neural re-ranking model. In the following, we describe how we learn contextualized term representations (Section \ref{sec:tk_context}) and how we transparently score their interactions (Section \ref{sec:tk_scoring}). Figure \ref{fig:tk_model_architecture} gives an overview of TK's architecture.

\subsection{Contextualized Term Representation}
\label{sec:tk_context}

TK uses a hybrid contextualization approach. The base representations are single-vector-per-word embeddings \cite{pennington2014glove}. We chose a simple word embedding structure over more complex methods -- such as FastText \cite{bojanowski2017enriching} or ELMo \cite{Peters2018} -- as it offers the following benefits in practice: Word embeddings are easy to pre-train on domain specific data \cite{Hofstaetter2019_ecir}; they require only one id per term, making the index consume less disk space, once prepared for re-ranking; most importantly, at query time, their selection is a fast memory lookup.

In the contextualization phase of the TK model, we process query ${q}_{1:m}$ and document sequences ${d}_{1:n}$ separately, however the learned parameters are shared. The input consists of two sequences of query and document ids. We employ the lookup based word embedding to select non-contextualized representations for each term. The hybrid-contextualized representation $\hat{t}_i$ of a term with word embedding $t_i$ over its whole input sequence $t_{1:n}$ is defined as:
\begin{equation}
\hat{t}_i = t_i * \alpha + \operatorname{context}(t_{1:n})_i * (1 - \alpha)
\end{equation}

We regulate the influence of the contextualization by the end-to-end learned $\alpha$ parameter. This allows the model to decide the intensity of the contextualization. We calculate the $\operatorname{context}(t_{1:n})$ with a set of Transformer layers \cite{vaswani2017attention}. First, the input sequence is fused with a positional encoding to form $p_{1:n}$, followed by a set of $l$ Transformer layers:  
\begin{equation}
\operatorname{Transformer_l}(p_{1:n}) = \operatorname{MultiHead}(\operatorname{FF}(p_{1:n})) + \operatorname{FF}(p_{1:n})
\end{equation}

Here, $\operatorname{FF}$ is a two-layer fully connected feed-forward layer including a non-linear activation function. The $MultiHead$ module projects the input sequence (stored as a matrix) to query, key, and value inputs of the scaled dot-product attention for each attention head. Then the results of the attention heads are concatenated and projected to the output:  
\begin{equation}
\begin{aligned} 
\operatorname{MultiHead}(p_{1:n}) &=\operatorname{Concat}(head_{1}, ..., head_{h}) W^{O} \\
\text{where } head_i &= \operatorname{softmax}\left(\frac{(p_{1:n} W_{i}^{Q}) (p_{1:n} W_{i}^{K})^{T}}{\sqrt{d_{k}}}\right) (p_{1:n} W_{i}^{V})
\end{aligned}
\end{equation}

We select Transformers for contextualization, because their positional encoding and sequence wide self-attention allows for local and global contextualization at the same time. This makes TK more powerful than previous local-only contextualization methods used in CONV-KNRM \cite{Dai2018} and CO-PACRR \cite{hui2018co}.

\subsection{Interaction Scoring}
\label{sec:tk_scoring}

After the contextualization, we match the query sequence $\hat{q}_{1:m}$ and document sequence $\hat{d}_{1:n}$ together in a single match-matrix $M~\in~\mathbb{R}^{q_{\text{len}} \times  d_{\text{len}}}$ with pairwise cosine similarity as interaction extractor:
\begin{equation}
M_{i,j} = \cos(\hat{q_i},\hat{d_j})
\end{equation}

Then, we transform each entry in $M$ with a set of $k$ RBF-kernels \cite{Xiong2017}. Each kernel focuses on a specific similarity range with center $\mu_{k}$. The size of all ranges is guided by $\sigma$. In contrast to Xiong et al. \cite{Xiong2017} we do not employ an exact match kernel -- as contextualized representation are not producing exact matches. Each kernel results in a matrix $K \in \mathbb{R}^{q_{\text{len}} \times  d_{\text{len}}}$: 
\begin{equation}
K^{k}_{i,j} = \exp \left(-\frac{\left(M_{i j}-\mu_{k}\right)^{2}}{2 \sigma^{2}}\right)
\end{equation}

Now, we process each kernel matrix in parallel, and we begin by summing the document dimension $j$ for each query term and kernel:
\begin{equation}
K^{k}_{i} = \sum_{j} K^{k}_{i,j}
\end{equation}

At this point -- as shown in Figure \ref{fig:tk_model_architecture} -- the model flow splits into two paths: log normalization and length normalization. The log normalization applies a logarithm with base $b$ to each query term before summing them up:
\begin{equation}
s^{k}_{\text{log}} = \sum_{i} \log_b\left( K^{k}_{i} \right)
\end{equation}

We enhance the pooling process with document length normalization. We dampen the magnitude of each query term signal by the document length:
\begin{equation}
s^{k}_{\text{len}} = \sum_{i} \frac{K^{k}_{i}}{d_{\text{len}}}
\end{equation}

Now, each kernel is represented by a single scalar, which is weighted with a simple linear layer to produce a scalar, for both the log-normalized and length normalized kernels: 
\begin{align}
s_{\text{log}} = s^{k}_{\text{log}} W_{1} && s_{\text{len}} = s^{k}_{\text{len}} W_{2}
\end{align}

Finally, we compute the final score of the query-document pair as a weighted sum of the log-normalized and the length-normalized scores:
\begin{equation}
s = s_{\text{log}} * \beta + s_{\text{len}} * \gamma 
\end{equation}

We employ kernel-pooling, because it makes inspecting temporary scoring results more feasible compared to pattern based scoring methods. Each kernel is applied to the full document. The row-wise and the column-wise summing of the match-matrix allow to inspect individual matches independent from each other. 
\section{TREC Deep Learning Track}
\label{sec:results}

Now, we describe the details of our experimentation pipeline (Section \ref{sec:experiment_setup}), configuration settings of our different runs (Section \ref{sec:runs}) and the results of the TREC annotations (Section \ref{sec:result_data}).

\subsection{Experiment Setup}
\label{sec:experiment_setup}

Our experimentation pipeline has two parts: 1) A first stage index and retrieval and 2) neural re-ranking model training and inference. For the first part we use the Anserini toolkit \cite{Yang2017} to compute the initial ranking lists, which we use to generate training and evaluation inputs for the neural models (only for the \textit{full} task). As basis for our neural model library we use PyTorch~\cite{pytorch2017} and AllenNLP~\cite{Gardner2017AllenNLP}. We tokenize the text with the fast BlingFire library\footnote{https://github.com/microsoft/BlingFire}. We train all neural models with a pairwise hinge loss. We use pre-trained GloVe~\cite{pennington2014glove} word embeddings with 300 dimensions\footnote{42B CommonCrawl lower-cased: \textit{https://nlp.stanford.edu/projects/glove/}}.

We cap the query length at $30$ tokens and the document length at $200$ tokens. For MSMARCO-Passage this only removes a modest amount of outliers, however, for the MSMARCO-Document collection a majority of documents is longer than 200 tokens. Increasing the cap to fully include most documents would render all evaluated neural IR models less effective. Only the TUW19-d3 model uses a larger cutoff of 800. We use the Adam \cite{kingma2014adam} optimizer with a learning rate of $10^{-4}$ for word embeddings and contextualization layers, $10^{-3}$ for all other network layers. We employ early stopping, based on the best MRR@10 value of the validation set. We use a training batch size of 64. We use a vocabulary of all terms with a minimum collection occurrence of 5. Regarding model-specific parameters, for the Transformer layers in TK we use 2 layers, each with 16 attention heads with size 32 and a feed-forward dimension of 100. For log-normalization in TK we use a base of 2. For kernel-pooling we set the number of kernels to $11$ with the mean values of the Gaussian kernels varying from $-1$ to $+1$ and standard deviation of $0.1$ for all kernels.

\subsection{Run overview}
\label{sec:runs}

Our runs are described in Table \ref{tab:submitted_runs}. We used the same model instance for both the evaluation of the \textit{re-ranking} task and the \textit{full ranking} task. For the \textit{full} task we generated initial rankings with Anserini using BM25 and utilized the validation sets to tune the re-ranking depth. For the \textit{re-ranking} task we used the entire provided initial ranking list. 

The \textit{TUW19-d3} model is the only submitted run diverging from the TK model description in Section \ref{sec:model}. It caps documents at 800 tokens and contextualizes the sequences in one block. Then, after the kernel-transformation of the cosine interactions is computed, we apply multiple pooling windows of different window sizes to the soft-histogram features. After that, we sort the window-scores and weigh the position independent sorted scores to form the final score.  

\begin{table}[t!]
    \centering
    \caption{Summary of our submitted TK runs}
    \label{tab:submitted_runs}
    \vspace{-0.3cm}
    \setlength\tabcolsep{5pt}
    \begin{tabular}{lL{6cm}}
       \toprule
       \textbf{Run}& \textbf{Description}\\
        \midrule
        \multicolumn{2}{c}{\textbf{Passages}} \\
        \textbf{TUW19-p1}  &  Using GloVe pre-trained (with min. 5 occurrence threshold), best validation run of multiple inits \\
        \textbf{TUW19-p2} &  Using FastText vectors instead of a simple word embedding \\
        \textbf{TUW19-p3} &  Ensemble of multiple TUW19-p1 configurations\\
        \midrule
        \multicolumn{2}{c}{\textbf{Documents}} \\
        \textbf{TUW19-d1}  &  Using GloVe and document training data\\
        \textbf{TUW19-d2} &  Using GloVe and passage training data \\
        \textbf{TUW19-d3} &  Using FasText embeddings \& windowed-kernel-pooling of different sizes, final score based on sorted window scores\\
        \bottomrule
    \end{tabular}
\end{table}

\subsection{Results}
\label{sec:result_data}

Now, we present our results for our validation set (sparsely labeled MSMARCO-DEV set) and the TREC judgements in Table \ref{tab:all_results}. Additionally, we highlight qualitative examples of the best and worst queries for two runs in Tables \ref{tab:passage_examples} and \ref{tab:doc_examples}. 

\begin{table*}[t!]
    \centering
    \caption{Evaluation results of our runs for the passage and document tasks.}
    \label{tab:all_results}
    \vspace{-0.3cm}
    \setlength\tabcolsep{5pt}
    \begin{tabular}{cl!{\color{lightgray}\vrule}rrr!{\color{lightgray}\vrule}rrr!{\color{lightgray}\vrule}rrr}
       \toprule
       &\multirow{2}{*}{\textbf{Run}}& 
       \multicolumn{3}{c!{\color{lightgray}\vrule}}{\textbf{MSMARCO-DEV}}&
       \multicolumn{3}{c!{\color{lightgray}\vrule}}{\textbf{TREC2019-Full}}&
       \multicolumn{3}{c}{\textbf{TREC2019-ReRank}}\\
       && MAP & nDCG & MRR@10 & MAP & nDCG & P@10 & MAP & nDCG & P@10\\
        \midrule
        \parbox[t]{2mm}{\multirow{3}{*}{\rotatebox[origin=c]{90}{\small\textbf{Passage}}}} &
        \textbf{TUW19-p1}   & 0.314 & 0.366 & 0.307 & 0.413 & 0.667 & 0.574 & 0.407 & 0.640 & 0.570 \\
        &\textbf{TUW19-p2}   & 0.316 & 0.369 & 0.310 & 0.416 & 0.671 & 0.577 & 0.396 & 0.636 & 0.565 \\
        &\textbf{TUW19-p3}   & \textbf{0.333} & \textbf{0.386} & \textbf{0.328} &\textbf{ 0.420} & \textbf{0.671} & \textbf{0.598} & \textbf{0.411} & \textbf{0.641} & \textbf{0.577}\\
        \midrule
        \parbox[t]{2mm}{\multirow{3}{*}{\rotatebox[origin=c]{90}{\small\textbf{Doc.}}}} &
        \textbf{TUW19-d1}   & 0.311 & 0.366 & 0.306 & 0.165 & 0.314 & 0.626 & 0.252 & 0.445 & 0.688 \\
        &\textbf{TUW19-d2}  & 0.312 & 0.365 & 0.303 & \textbf{0.205} & \textbf{0.382} & \textbf{0.633} & 0.239 & 0.445 & 0.681 \\
        &\textbf{TUW19-d3}  & \textbf{0.314} & \textbf{0.369} & \textbf{0.309} & 0.184 & 0.333 & 0.626 & \textbf{0.271} & \textbf{0.465} & \textbf{0.730} \\
        \bottomrule
    \end{tabular}
\end{table*}

\subsubsection{Passage Task}

For the passage task our results show that different configurations of TK have similar results. Especially, the difference between a GloVe embedding with a minimum threshold of 5 (p1) and a FastText embedding without out-of-vocabulary (OOV) terms (p2) is marginal. We assume this is due to the ability of the contextualization to overcome OOV and infrequent terms, which have been shown to negatively impact simpler neural models \cite{Hofstaetter2019_sigir}. Ensembling a model (p3), does provide some benefit, however, the difference is stronger in the loosely judged DEV set and smaller in the thoroughly judged TREC annotations.

In addition to the evaluation metrics, we selected the best and worst performing queries from the TREC'19 set and show them in Table \ref{tab:passage_examples}. Due to the small amount of queries evaluated, there is no clear distinction in the types of information needs that perform better or worse. 

\begin{table}[t!]
    \centering
    \caption{Best \& worst queries for TUW19-p3 full}
    \label{tab:passage_examples}
    \vspace{-0.3cm}
    \setlength\tabcolsep{5pt}
    \begin{tabular}{llL{5.5cm}}
       \toprule
       \multicolumn{3}{c}{\textbf{Best}}\\
       \textbf{Id} & \textbf{AP} & \textbf{Query Text}\\
        \midrule
        146187 & 0.851  &  difference between a mcdouble and a double cheeseburger \\
        156493 & 0.8225 &  do goldfish grow \\ 
        168216 & 0.9495 &  does legionella pneumophila cause pneumonia \\
        359349 & 0.788  &  how to find the midsegment of a trapezoid \\
        855410 & 1      &  what is theraderm used for \\ 
        \toprule
        \multicolumn{3}{c}{\textbf{Worst}}\\
        \textbf{Id} & \textbf{AP} & \textbf{Query Text}\\
        \midrule
        1063750 & 0.015   &  why did the us volunterilay enter ww1 \\
        1110199 & 0.0911  &  what is wifi vs bluetooth \\
        1112341 & 0.1056  &  what is the daily life of thai people \\
        1113437 & 0.095   &  what is physical description of spruce \\
        19335   & 0       &  anthropological definition of environment \\
        207786  & 0.0919  &  how are some sharks warm blooded \\
        489204  & 0.0583  &  right pelvic pain causes \\
        \bottomrule
    \end{tabular}
    \vspace{-0.2cm}
\end{table}

\subsubsection{Document Task}

For the document ranking the different configurations and network structures of TK seem very similar when looking at the DEV set in Table \ref{tab:all_results}. However, the TREC annotations reveal large differences between the full and re-Ranking task as well as the submitted configurations. 

We assume that a major factor for the differences between \textit{full} and \textit{re-ranking} tasks is that for the \textit{full} task we tuned the re-ranking depth on the MRR@10 score of the validation set as proposed by Hofst\"{a}tter et al. \cite{Hofstaetter2019_sigir}. For the passage models we found the maximum evaluated depth (1000 documents per query) to be the best, but for the document task the tuned threshold (on the MSMARCO-DEV set) is much lower than the 100 documents of the full ranking task. The re-ranking depths are 29 for \textit{TUW19-d1}, 60 for \textit{TUW19-d2} and 31 for \textit{TUW19-d3}. For the sparsely judged DEV set this brings improvements, however for the thoroughly judged TREC queries we decrease the effectiveness substantially between 31 and 100 re-ranked documents. This shows the importance of evaluating thoroughly judged queries and the need to revisit the threshold parameter for the new dataset. 

The best performing TK model (d3) is document-specific with windowed kernel-pooling and tuned re-ranking depth. This result shows that the pure passage ranking TK model (d1,d2) is unsuited for documents and we explored a first strategy on handling longer text, still we believe there is more potential for special network architectures for documents in the future.

Following the passage result, we also highlight the best and worst performing queries on the document task for the d3 run and the re-ranking task in Table \ref{tab:doc_examples}.

\begin{table}[t!]
    \centering
    \caption{Best \& worst queries for TUW19-d3 re-rank}
    \label{tab:doc_examples}
    \vspace{-0.3cm}
    \setlength\tabcolsep{5pt}
    \begin{tabular}{llL{5.5cm}}
       \toprule
       \multicolumn{3}{c}{\textbf{Best}}\\
       \textbf{Id} & \textbf{AP} & \textbf{Query Text}\\
        \midrule

        287683 & 0.7917 & how many liberty ships were built in brunswick \\
        405717 & 0.4675 & is cdg airport in main paris \\
        855410 & 1      & what is theraderm used for \\
        962179 & 0.7067 & when was the salvation army founded \\

        \toprule
        \multicolumn{3}{c}{\textbf{Worst}}\\
        \textbf{Id} & \textbf{AP} & \textbf{Query Text}\\
        \midrule

        1037798 & 0.0076 & who is robert gray \\ 
        1063750 & 0.0389 & why did the us volunterilay enter ww1\\ 
        1106007 & 0.0414 & define visceral?\\ 
        1112341 & 0.0277 & what is the daily life of thai people\\ 
        
        443396 & 0.0316 & lps laws definition \\
        451602 & 0.0276 & medicare's definition of mechanical ventilation \\
        47923  & 0.0625 & axon terminals or synaptic knob definition \\
        489204 & 0.0736 & right pelvic pain causes \\

        \bottomrule
    \end{tabular}
    \vspace{-0.2cm}
\end{table}

\begin{figure*}[!t]
    \centering
    \noindent
    \begin{minipage}{\textwidth}
    \begin{center}
        {\color[RGB]{76, 76, 90}Query (Id:2)} \textbf{androgen receptor define}
    \end{center}
    \vspace{0.2cm}
    \end{minipage} 
    \begin{minipage}{.5\textwidth}
    \centering
      Rank: TK \raisebox{.5pt}{\textcircled{\raisebox{-.85pt} {1}}}, BM25 \raisebox{.5pt}{\textcircled{\raisebox{-.85pt} {9}}} \textbf{(judged as relevant}, Id: 4339068) 
      \newline
    \end{minipage}
    \begin{minipage}{.5\textwidth}
    \centering
      Rank: TK \raisebox{.5pt}{\textcircled{\raisebox{-.85pt} {8}}}, BM25 \raisebox{.5pt}{\textcircled{\raisebox{-.85pt} {1}}} (not relevant, Id: 1782337)
      \newline
    \end{minipage}
    \begin{minipage}[t]{.35\textwidth}%
    \vspace{-0.21cm}
{\color[RGB]{0,151,20}\uline{The}} {\color[RGB]{0,151,20}\uline{androgen}} {\color[RGB]{190,60,60}\uuline{receptor}} {\color[RGB]{0,151,20}\uline{(}} {\color[RGB]{0,151,20}\uline{AR}} {\color[RGB]{0,151,20}\uline{)}} {\color[RGB]{0,151,20}\uline{,}} {\color[RGB]{0,151,20}\uline{also}} {\color[RGB]{0,151,20}\uline{known}} {\color[RGB]{0,151,20}\uline{as}} {\color[RGB]{0,151,20}\uline{NR3C4}} {\color[RGB]{0,151,20}\uline{(}} {\color[RGB]{145,145,149}nuclear} {\color[RGB]{0,151,20}\uline{receptor}} {\color[RGB]{0,151,20}\uline{subfamily}} {\color[RGB]{145,145,149}3} {\color[RGB]{145,145,149},} {\color[RGB]{145,145,149}group} {\color[RGB]{145,145,149}C} {\color[RGB]{145,145,149},} {\color[RGB]{145,145,149}member} {\color[RGB]{145,145,149}4} {\color[RGB]{0,151,20}\uline{)}} {\color[RGB]{145,145,149},} {\color[RGB]{0,151,20}\uline{is}} {\color[RGB]{0,151,20}\uline{a}} {\color[RGB]{0,151,20}\uline{type}} {\color[RGB]{145,145,149}of} {\color[RGB]{145,145,149}nuclear} {\color[RGB]{0,151,20}\uline{receptor}} {\color[RGB]{145,145,149}that} {\color[RGB]{0,151,20}\uline{is}} {\color[RGB]{145,145,149}activated} {\color[RGB]{145,145,149}by} {\color[RGB]{145,145,149}binding} {\color[RGB]{145,145,149}either} {\color[RGB]{145,145,149}of} {\color[RGB]{145,145,149}the} {\color[RGB]{145,145,149}androgenic} {\color[RGB]{145,145,149}hormones} {\color[RGB]{145,145,149},} {\color[RGB]{145,145,149}testosterone} {\color[RGB]{145,145,149},} {\color[RGB]{0,151,20}\uline{or}} {\color[RGB]{145,145,149}dihydrotestosterone} {\color[RGB]{145,145,149}in} {\color[RGB]{145,145,149}the} {\color[RGB]{145,145,149}cytoplasm} {\color[RGB]{145,145,149}and} {\color[RGB]{145,145,149}then} {\color[RGB]{145,145,149}translocating} {\color[RGB]{0,151,20}\uline{into}} {\color[RGB]{145,145,149}the} {\color[RGB]{145,145,149}nucleus} {\color[RGB]{145,145,149}.} {\color[RGB]{0,151,20}\uline{in}} {\color[RGB]{145,145,149}some} {\color[RGB]{145,145,149}cell} {\color[RGB]{145,145,149}types} {\color[RGB]{145,145,149},} {\color[RGB]{145,145,149}testosterone} {\color[RGB]{145,145,149}interacts} {\color[RGB]{145,145,149}directly} {\color[RGB]{145,145,149}with} {\color[RGB]{145,145,149}androgen} {\color[RGB]{0,151,20}\uline{receptors}} {\color[RGB]{145,145,149},} {\color[RGB]{145,145,149}whereas} {\color[RGB]{145,145,149},} {\color[RGB]{145,145,149}in} {\color[RGB]{145,145,149}others} {\color[RGB]{145,145,149},} {\color[RGB]{145,145,149}testosterone} {\color[RGB]{145,145,149}is} {\color[RGB]{145,145,149}converted} {\color[RGB]{145,145,149}by} {\color[RGB]{145,145,149}5} {\color[RGB]{145,145,149}-} {\color[RGB]{145,145,149}alpha} {\color[RGB]{145,145,149}-} {\color[RGB]{145,145,149}reductase} {\color[RGB]{145,145,149}to} {\color[RGB]{145,145,149}dihydrotestosterone} {\color[RGB]{145,145,149},} {\color[RGB]{145,145,149}an} {\color[RGB]{145,145,149}even} {\color[RGB]{145,145,149}more} {\color[RGB]{145,145,149}potent} {\color[RGB]{145,145,149}agonist} {\color[RGB]{145,145,149}for} {\color[RGB]{145,145,149}androgen} {\color[RGB]{0,151,20}\uline{receptor}} {\color[RGB]{145,145,149}activation} {\color[RGB]{145,145,149}.}
\end{minipage}
    \hfill\begin{minipage}[t]{.14\textwidth}%
    \centering
    \begin{tabular}[t]{lr}
         \textbf{$\boldsymbol\mu_{k}$} & \textbf{$\boldsymbol{s^{k}_{\text{log}}}$} \\
         1 & -3.1 \\
         {\color[RGB]{202, 70, 70}\uuline{0.9} }& {\color[RGB]{202, 70, 70}-0.6} \\
         {\color[RGB]{0,151,20}\uline{0.7}} & {\color[RGB]{0,151,20}2.3} \\
         0.5 & -1.6 \\
         0.3 & -3.3 \\
         Rest & -14.6\\
         \midrule
         \textbf{$\boldsymbol{s_{\text{log}}}$} & -11.6\\
         \textbf{$\boldsymbol{s_{\text{len}}}$} & 1.1\\
         \midrule
         $\boldsymbol{s}$ & -10.5\\

    \end{tabular}
    \end{minipage}\hfill\vline\hfill
    \begin{minipage}[t]{.14\textwidth}%
    \centering
      \begin{tabular}[t]{lr}
         \textbf{$\boldsymbol\mu_{k}$} & \textbf{$\boldsymbol{s^{k}_{\text{log}}}$} \\
         1 & -5.0\\
         {\color[RGB]{202, 70, 70}\uuline{0.9}} & {\color[RGB]{202, 70, 70}-1.5}\\
         {\color[RGB]{0,151,20}\uline{0.7}} & {\color[RGB]{0,151,20} 1.9}\\
         0.5 & -1.3\\
         0.3 & -2.3\\
         Rest & -14.6\\
         \midrule
         \textbf{$\boldsymbol{s_{\text{log}}}$} & -12.8\\
         \textbf{$\boldsymbol{s_{\text{len}}}$} & 0.9\\
         \midrule
         $\boldsymbol{s}$ & -11.9\\
    \end{tabular}
    \end{minipage}
    \hfill\begin{minipage}[t]{.35\textwidth}%
    \vspace{-0.21cm}
{\color[RGB]{145,145,149}Enzalutamide} {\color[RGB]{0,151,20}\uline{is}} {\color[RGB]{0,151,20}\uline{an}} {\color[RGB]{0,151,20}\uline{androgen}} {\color[RGB]{0,151,20}\uline{receptor}} {\color[RGB]{145,145,149}inhibitor} {\color[RGB]{0,151,20}\uline{that}} {\color[RGB]{0,151,20}\uline{acts}} {\color[RGB]{145,145,149}on} {\color[RGB]{145,145,149}different} {\color[RGB]{145,145,149}steps} {\color[RGB]{145,145,149}in} {\color[RGB]{0,151,20}\uline{the}} {\color[RGB]{145,145,149}androgen} {\color[RGB]{0,151,20}\uline{receptor}} {\color[RGB]{145,145,149}signaling} {\color[RGB]{145,145,149}pathway} {\color[RGB]{145,145,149}.} {\color[RGB]{145,145,149}Enzalutamide} {\color[RGB]{145,145,149}has} {\color[RGB]{145,145,149}been} {\color[RGB]{0,151,20}\uline{shown}} {\color[RGB]{0,151,20}\uline{to}} {\color[RGB]{145,145,149}competitively} {\color[RGB]{145,145,149}inhibit} {\color[RGB]{145,145,149}androgen} {\color[RGB]{0,151,20}\uline{binding}} {\color[RGB]{0,151,20}\uline{to}} {\color[RGB]{145,145,149}androgen} {\color[RGB]{0,151,20}\uline{receptors}} {\color[RGB]{145,145,149}and} {\color[RGB]{145,145,149}inhibit} {\color[RGB]{145,145,149}androgen} {\color[RGB]{0,151,20}\uline{receptor}} {\color[RGB]{145,145,149}nuclear} {\color[RGB]{145,145,149}translocation} {\color[RGB]{145,145,149}and} {\color[RGB]{145,145,149}interaction} {\color[RGB]{145,145,149}with} {\color[RGB]{145,145,149}DNA} {\color[RGB]{145,145,149}.}
\end{minipage}  
    \caption{TK's scoring results of two MSMARCO-Passage documents: We highlight two of the most distinct kernels -- both indicating close similarities (0.9 \& 0.7). In the text words are colored and underlined if they are closest to the center of the highlighted kernel. Individual kernel results (model weights included) are displayed in the middle for each document.}    \label{fig:interpretability_2col}
    \vspace{-0.2cm}
\end{figure*}

\section{Interpretability}
\label{sec:interpretability}

We now highlight the interpretation capabilities of the TK model with a qualitative example from the MSMARCO-Passage-DEV set. We focus on the following scenario: a user would like to know why the neural model replaced the first result (a non-relevant document) of the first stage ranking with an actual relevant document. For this, we offer a side-by-side comparison view of two documents. 

Figure \ref{fig:interpretability_2col} shows the comparison of two documents for the query \textit{"androgen receptor define"}. On the left side is a document judged as relevant, which is placed on the first position by TK. On the right side is the formerly first document (as determined by BM25), which is not the correct answer and only partially relevant to the query -- TK moved it to a lower position. 

We show each document with its full-text and a selection of temporary results of TK. We aim to identify and highlight the differences that result in different ranking scores. We color words according to their closest affiliation with a kernel. An important fact to consider is the soft-matching nature of the kernels: A term is counted in more than one kernel at a time. For example, this explains the difference in kernel $\mu=1$, even though no word is more closely associated with that kernel and therefore we omitted a color.

From the highlighted kernel scores  (${s^{k}_{\text{log}}}$) it is apparent that the left document has more stronger matches than the right one, leading to higher scores. If we look at the corresponding colored words we observe that the sentence containing the definition in the left is most relevant to the query: \textit{{\color[RGB]{0,151,20}\uline{The}} {\color[RGB]{0,151,20}\uline{androgen}} {\color[RGB]{190,60,60}\uuline{receptor}} {\color[RGB]{0,151,20}\uline{(}} {\color[RGB]{0,151,20}\uline{AR}} {\color[RGB]{0,151,20}\uline{)}} {\color[RGB]{0,151,20}\uline{,}} {\color[RGB]{0,151,20}\uline{also}} {\color[RGB]{0,151,20}\uline{known}} {\color[RGB]{0,151,20}\uline{as}} {\color[RGB]{0,151,20}\uline{NR3C4}} {\color[RGB]{0,151,20}\uline{(}} {\color[RGB]{145,145,149}nuclear} {\color[RGB]{0,151,20}\uline{receptor}} {\color[RGB]{0,151,20}\uline{subfamily}}}. Even though TK does not contain a mechanism for strictly categorizing a region as relevant, it does so indirectly by strongly matching most terms in this region. Of particular interest to us is the fact that the contextualization of TK learns to match the query term \textit{"define"} with words and phrases that make up a definition: \textit{"also known as"}, \textit{"subfamily"}, \textit{"is a type"} as well as the parentheses. This exceeds simple synonym mapping, suggesting once more the importance of training contextualized and relevance specific encoding models. 

This analysis demonstrates the potential for future work on keyword based search. When a collection is not queried with natural language questions, but only keywords, one could expand such keyword queries with terms like \textit{"definition"} or \textit{"meaning"} both during training and inference of neural models, to promote documents closer related to the core of the information need. 
We are aware that our approach does not enable full interpretability. We do not look deeper than the pairwise similarity values, and we currently cannot explain why certain words are similar and why some are not. Additionally, an interactive version as part of a search result page would allow more flexibility to explore different kernels and query terms. Nevertheless, we view this approach as a first step to open up the black-box of the neural re-ranking model.
\vspace{-0.2cm}
\section{Conclusion}

Our aim in the TREC 2019 Deep Learning track was to evaluate a neural re-ranking model, which balances efficiency, effectiveness, and interpretability. We submitted results of our TK (Transformer-Kernel) model: a neural re-ranking model for ad-hoc search using an efficient contextualization mechanism. For the passage task our results show that different configurations of TK lead to similar results. Ensembling a model does provide some benefit, however the difference is stronger in the loosely judged DEV set and smaller in the thoroughly judged TREC annotations. For the document ranking the different configurations and network structures of TK seem very similar when looking at the DEV set. The TREC annotations however reveal large differences between the full and re-ranking task and the submitted configurations. The best performing TK model is document-length-specific with windowed kernel-pooling and tuned re-ranking depth. This shows the potential for specialized architectures for neural document re-ranking.
\vspace{-0.1cm}

\bibliographystyle{ACM-Reference-Format}
\bibliography{my-references}

\end{document}